\documentclass[prb,twocolumn]{revtex4}

\usepackage{graphicx}

\begin{document}

\title{
Magnetoelectric effects in heavy-fermion superconductors without
inversion symmetry 
}

\author{Satoshi Fujimoto} 
\affiliation{
Department of Physics,
Kyoto University, Kyoto 606-8502, Japan
}

\date{\today}

\begin{abstract}
We investigate effects of strong electron correlation on
magnetoelectric transport phenomena 
in noncentrosymmetric superconductors 
with particular emphasis on its application to
the recently discovered heavy-fermion superconductor CePt$_3$Si.
Taking into account electron correlation effects in a formally exact way,
we obtain the expression of the magnetoelectric coefficient
for the Zeeman-field-induced paramagnetic supercurrent, of which
the existence was predicted more than a decade ago.
It is found that in contrast to the usual Meissner current, which is
much reduced by the mass renormalization factor 
in the heavy-fermion state, the paramagnetic supercurrent is not affected by 
the Fermi liquid effect.
This result implies that the experimental observation of
the magnetoelectric effect is more feasible in heavy-fermion systems 
than that in conventional metals with moderate effective mass.
\end{abstract}


\maketitle
\section{Introduction}

It has been discussed for decades that metallic 
systems with noncentrosymmetric crystal structure
may exhibit non-trivial magnetoelectric 
effects.~\cite{lev,ede0,ede1,ede2,yip}
The existence of an asymmetric potential gradient $\vec{\nabla}V$ due to
the noncentrosymmetric structure gives rise to
the spin-orbit interaction 
$(\vec{p}\times \vec{\nabla}V)\cdot\vec{\sigma}$,
which breaks the individual inversion and spin rotation symmetry.
As a result, the charge or energy current operator 
may couple to the spin density operator.
In this context, current flows induced by an applied magnetic field, and
current-flow-driven magnetization 
have been investigated extensively
both in normal metals~\cite{lev,ede0} and 
superconductors.~\cite{ede1,ede2,yip}
In particular, Edelstein predicted the remarkable magnetoelectric effect 
in superconducting states; i.e.
in noncentrosymmetric superconductors,
the Zeeman field induces a supercurrent, and vice versa,
the supercurrent flow induces a magnetization.~\cite{ede1,ede2}
Later, the former effect is elegantly re-formulated by Yip in terms of
the ``van Vleck'' contribution which stems 
from the inversion-symmetry-breaking spin-orbit interaction.~\cite{yip} 
Since a static magnetic field can not induce dissipative current flows,
the Zeeman-energy-induced current should vanish in the normal state.
However, in the superconducting state, 
the existence of the paramagnetic supercurrent
is not forbidden in the absence of the inversion symmetry.
The recent discovery of superconducting materials without inversion
symmetry such as CePt$_3$Si, UIr, and Cd$_2$Re$_2$O$_7$
stimulates the renewed interest 
in this issue.~\cite{bau,uir,cd,take,hari,met,yogi}
Under an applied magnetic field, the Meissner diamagnetic supercurrent
in addition to the Zeeman-field-induced paramagnetic supercurrent
should exist. Thus, it is important for the experimental observation
of this effect to discriminate between these two supercurrents.

In the present paper, we would like to investigate
the Fermi liquid corrections to this Edelstein's magnetoelectric effect,
which may be important for the application to heavy-fermion superconductors
such as CePt$_3$Si and UIr.
We obtain the formula for the magnetoelectric effect coefficient
taking into account Fermi liquid corrections exactly.
The most important finding is that
the Zeeman-energy-induced paramagnetic supercurrent is not at all
affected by electron correlation effects provided that
ferromagnetic spin fluctuation is not developed, in contrast with
the diamagnetic Meissner current of which the magnitude is
much reduced by the mass renormalization effect. 
This result implies that the experimental detection of
the paramagnetic supercurrent in heavy-fermion superconductors may be
more feasible than that in weakly correlated metals.

The organization of this paper is as follows.
In Sec.II, we present the basic formulation of the Fermi liquid theory
for a model system without inversion symmetry.
We would like to make a brief comment on the superconducting state realized in
CePt$_3$Si in Sec.III.
In Sec.IV, the exact formula of the magnetoelectric coefficient
is obtained. In Sec.V, the implication for
the experimental observation of this effect is discussed.
Summary and discussion are given in the last section.

\section{Model and analysis based on the Fermi liquid theory}

As a simplest model which realizes the broken inversion symmetry,
we consider an interacting electron system with the Rashba spin-orbit
interaction.~\cite{ras}
The Hamiltonian is given by, 
\begin{eqnarray}
\mathcal{H}&=&\sum_{p,\sigma} \varepsilon_p c^{\dagger}_{\sigma p}c_{\sigma p}
+\alpha\sum_{p,\sigma\sigma'}
(\vec{p}\times \vec{n})\cdot\vec{\sigma}_{\sigma\sigma'}
c^{\dagger}_{\sigma p}c_{\sigma' p} \nonumber \\
&&+U\sum_i n_{\uparrow i}n_{\downarrow i}, \label{ham}
\end{eqnarray}
where $c^{\dagger}_{\sigma p}$ ($c_{\sigma p}$)
is the creation (annihilation) operator for an electron with spin $\sigma$
and momentum $p$.  The number density operator at 
the site $i$, $n_{\sigma i}=c^{\dagger}_{\sigma i}c_{\sigma i}$.
The second term of (\ref{ham}) is the Rashba spin-orbit interaction
which incorporates the broken inversion symmetry. 
Here, 
the unit vector parallel to the asymmetric potential gradient is given by
$\vec{n}=(0,0,1)$.
This system is considered to be a model of CePt$_3$Si, with which we are
mainly concerned in this paper.
The $f$-electron of CePt$_3$Si is in the $\Gamma_7$ Kramers 
doublet state.~\cite{met}
Expanding the $\Gamma_7$ doublet in terms of the $s_z=1/2$ and $-1/2$ basis,
we found that the Rashba spin-orbit interaction term expressed 
in term of the $\Gamma_7$ basis has the matrix structure
given above up to a constant factor which can be absorbed into
the re-definition of the coupling constant $\alpha$. 
Thus, in the case of CePt$_3$Si, the spin index $\sigma$ in (\ref{ham})
represents the $\Gamma_7$ Kramers doublet.

In the following, we do not specify the pairing mechanism
of superconductivity, but assume that the superconducting state is realized
by an effective pairing interaction with an angular momentum $l\geq 1$
which may stem from the on-site Coulomb interaction in (\ref{ham}) or may have
any other origin not included in the Hamiltonian (\ref{ham}).
Then, we can analyze electron correlation effects on this superconducting state
in a formally exact way by using the superconducting 
Fermi liquid theory.~\cite{leg,mig,fuji}

In the conventional Nambu representation,\cite{sch}
the inverse of the single-particle
Green's function is defined as,
\begin{eqnarray}
\hat{\mathcal{G}}^{-1}(p)=
\left(
\begin{array}{cc}
   i\varepsilon_n-\hat{H}(p) & -\hat{\Delta}(p)   \\
   -\hat{\Delta}^{\dagger}(p) & i\varepsilon_n+\hat{H}^{t}(-p)
\end{array}
\right), \label{gin}
\end{eqnarray}
where $p=(\vec{p},i\varepsilon)$, and,
\begin{eqnarray}
\hat{H}(p)=\hat{H}_0(p)+\hat{\Sigma}(p), \label{h1}
\end{eqnarray}
\begin{eqnarray}
\hat{H}_0=\varepsilon_p-\mu+\alpha(\vec{p}\times\vec{n})\cdot\vec{\sigma}
-\mu_{\rm B}\sigma_xH_x, \label{h0}
\end{eqnarray}
with $\mu$ chemical potential.
Here, to discuss the magnetoelectric effect,
we take into account the Zeeman magnetic field
in the $x$ direction, $H_x$.
For simplicity, we assume that the $g$-value is equal to 2.
The self-energy matrix $\hat{\Sigma}$ consists of 
both diagonal and off-diagonal components,
\begin{eqnarray}
\hat{\Sigma}=
\left(
\begin{array}{cc}
 \Sigma_{\uparrow\uparrow}(p)   & \Sigma_{\uparrow\downarrow}(p)  \\
 \Sigma_{\downarrow\uparrow}(p) & \Sigma_{\downarrow\downarrow}(p)
\end{array}
\right). \label{self}
\end{eqnarray}
We can easily see from the symmetry argument that
under the applied in-plane magnetic field,
$\Sigma_{\uparrow\uparrow}(p)=\Sigma_{\downarrow\downarrow}(p)
\equiv\Sigma(p)$, and 
$\Sigma_{\downarrow\uparrow}(\vec{p},i\varepsilon)=
\Sigma_{\uparrow\downarrow}^{*}(\vec{p},-i\varepsilon)$. 
The superconducting gap function is
$\{\hat{\Delta}(p)\}_{\alpha\beta}=\Delta_{\alpha\beta}(p)$
($\alpha,\beta=\uparrow,\downarrow$).

$i\varepsilon_n-\hat{H}(p)$ and $i\varepsilon_n+\hat{H}^t(-p)$ in 
$\hat{\mathcal{G}}^{-1}(p)$
are diagonalized by the unitary transformation 
$\hat{\mathcal{A}}(p)\hat{\mathcal{G}}^{-1}(p)\hat{\mathcal{A}}^{\dagger}(p)$
with,
\begin{eqnarray}
\hat{\mathcal{A}}(p)=
\left(
\begin{array}{cc}
\hat{U}(\vec{p}) & 0 \\
 0 & \hat{U}^{t\dagger}(-\vec{p})
\end{array}
\right),
\end{eqnarray}
\begin{eqnarray}
\hat{U}(\vec{p})=
\frac{1}{\sqrt{2}}\left(
\begin{array}{cc}
1  &   i\hat{t}_{-} \\
i\hat{t}_{+} & 1 
\end{array}
\right)
\end{eqnarray}
where $\hat{t}_{\pm}=\hat{t}_x\pm i\hat{t}_y$, and  
$\hat{t}_x$, $\hat{t}_y$ are, respectively, the $x$ and $y$ components
of the unit vector $\vec{\hat{t}}_p\equiv\vec{t}_p/|\vec{t}_p|$ with
\begin{eqnarray}
\vec{t}_p(i\varepsilon)=
(p_x+\frac{1}{\alpha}{\rm Im}\Sigma_{\uparrow\downarrow},
p_y+\frac{1}{\alpha}{\rm Re}\Sigma_{\uparrow\downarrow}
-\frac{\mu_{\rm B}H_x}{\alpha},0). \label{tp}
\end{eqnarray}
As seen from eqs.(\ref{h1}),(\ref{h0}), and (\ref{self}),
the main effect of the off-diagonal self-energy 
$\Sigma_{\uparrow\downarrow}$ is to renormalize
the Rashba interaction term, replacing the momentum $\vec{p}$ 
in the Rashba term with the vector $\vec{t}_p$.
Since the on-site Coulomb interaction does not change 
the symmetry of the system, the off-diagonal self-energy
should satisfy the following condition in the absence of magnetic fields;
\begin{eqnarray}
{\rm Re}\Sigma_{\uparrow\downarrow}(p_x,-p_y)=
-{\rm Re}\Sigma_{\uparrow\downarrow}(p_x,p_y),
\end{eqnarray}
\begin{eqnarray}
{\rm Im}\Sigma_{\uparrow\downarrow}(-p_x,p_y)=
-{\rm Im}\Sigma_{\uparrow\downarrow}(p_x,p_y),
\end{eqnarray}
and ${\rm Re}\Sigma_{\uparrow\downarrow}$ 
(${\rm Im}\Sigma_{\uparrow\downarrow}$) is 
an even function of $p_x$ ($p_y$). 

In the normal state, 
the single-particle excitation energy $\varepsilon_{p\tau}^{*}$ 
for the quasi-particle with the helicity $\tau=\pm 1$ 
is given by the solution of the equation $z-\hat{H}(\vec{p},z)=0$,
which is, in the diagonalized representation,
\begin{equation}
\varepsilon_{p\tau}^{*}+\mu-\varepsilon_{p}
-\tau\alpha|\vec{t}_p(\varepsilon_{p\tau}^{*})|
-{\rm Re}\Sigma(\vec{p},\varepsilon_{p\tau}^{*})=0.
\end{equation}

The gap functions $\hat{\Delta}(p)$ and $\hat{\Delta}^{\dagger}(p)$ in
$\hat{\mathcal{G}}^{-1}(p)$ are also diagonalized by the 
unitary transformation 
$\hat{\mathcal{A}}(p)\hat{\mathcal{G}}^{-1}(p)\hat{\mathcal{A}}^{\dagger}(p)$
provided that the gap function has the following structure,
\begin{equation}
\hat{\Delta}(p)=\Delta_s(p)i\sigma_y
+\Delta_t(p)(\vec{\hat{t}}_p\times \vec{n})\cdot\vec{\sigma}i\sigma_y.
\label{gap}
\end{equation}
Here $\Delta_s(p)$ and $\Delta_t(p)$ are even functions of momentum $\vec{p}$.
This means that the spin singlet and triplet component is mixed 
in the diagonalized basis labeled by $\tau=\pm 1$, and
the $\vec{d}$ vector of the triplet component is 
$\vec{\hat{t}}_p\times \vec{n}$.~\cite{ede1,gor}
In the case that $\hat{\Delta}(p)$ is not expressed as eq.(\ref{gap}),
$\hat{H}(p)$ and $\hat{\Delta}(p)$ can not be diagonalized simultaneously,
and the non-zero off-diagonal components of $\hat{\Delta}(p)$ which
correspond to the Cooper pairing between the different Fermi surfaces
induce pair-breaking effects, resulting in the decrease of the transition
temperature $T_c$.
Thus, the highest transition temperature is achieved by
the gap function given by eq.(\ref{gap}).~\cite{fri} 
The realization of the gap function (\ref{gap}) 
in the case with no inversion center is
also elucidated by the group theoretical argument.~\cite{ser}

Taking the inverse of eq.(\ref{gin}), we have,
\begin{eqnarray}
\hat{\mathcal{G}}(p)=
\left(
\begin{array}{cc}
 \hat{G}(p)  & \hat{F}(p)   \\
  \hat{F}^{\dagger}(p) &  -\hat{G}^{t}(-p) 
\end{array}
\right), \label{g1}
\end{eqnarray}
where
\begin{eqnarray}
\hat{G}(p)=\sum_{\tau=\pm 1}\frac{1+\tau(\vec{\hat{t}}_p\times\vec{n})
\cdot\vec{\sigma}}{2}G_{\tau}(p),
\end{eqnarray}
\begin{eqnarray}
\hat{F}(p)=\sum_{\tau=\pm 1}\frac{1+\tau(\vec{\hat{t}}_p\times\vec{n})
\cdot\vec{\sigma}}{2}i\sigma_yF_{\tau}(p),
\end{eqnarray}
and,
\begin{eqnarray}
G_{\tau}(p)=\frac{z_{p\tau}(i\varepsilon+\varepsilon_{p\tau}^{*})}
{(i\varepsilon+i\gamma {\rm sgn}\varepsilon)^2-E^{2}_{p\tau}},
\end{eqnarray}
\begin{eqnarray}
F_{\tau}(p)=\frac{z_{p\tau}\Delta_{\tau}(p)}
{(i\varepsilon+i\gamma {\rm sgn}\varepsilon)^2-E^{2}_{p\tau}}.
\end{eqnarray}
Here the mass renormalization factor is,
\begin{eqnarray}
z_{p\tau}&=&\biggl[1-\frac{\partial {\rm Re}\Sigma(p)}
{\partial (i\varepsilon)} \nonumber \\
&+&\tau\left(
\hat{t}_x\frac{\partial {\rm Im}\Sigma_{\uparrow\downarrow}(p)}
{\partial (i\varepsilon)}
+\hat{t}_y\frac{\partial {\rm Re}\Sigma_{\uparrow\downarrow}(p)}
{\partial (i\varepsilon)}
\right)
\biggr]^{-1}\biggr|_{i\varepsilon=E_{p\tau}},
\end{eqnarray}
and $\gamma$ is the quasi-particle damping.
The single-particle excitation energy is
$E_{p\tau}=\sqrt{\varepsilon_{p\tau}^{* 2}+\Delta_{\tau}^2(p)}$
with $\Delta_{p\tau}=z_{p\tau}(\Delta_s(p)+\tau\Delta_t(p))$.

The superconducting gap function $\Delta_{\alpha\beta}$ and 
the transition temperature are determined by
the self-consistent gap equation,
\begin{eqnarray}
\Delta_{\alpha\beta}=T\sum_{n,p}{\rm Tr}[\hat{\Gamma}^{\alpha\beta}(p,p')
\hat{F}(p')],
\end{eqnarray}
where we have introduced the four-point vertex function matrix
$\{\hat{\Gamma}^{\alpha\beta}(p,p')\}_{\gamma\delta}$ which is diagrammatically
expressed as shown in FIG.1.
We expand the four-point vertex in the particle-particle channel
$\{\hat{\Gamma}^{\alpha\beta}(p,p')\}_{\gamma\delta}$
in terms of the basis of the irreducible representations of
the point group,
and consider a component 
$\{\hat{\Gamma}^{\alpha\beta}_a(p,p')\}_{\gamma\delta}$  which corresponds to
the pairing state giving highest $T_c$.
This pairing interaction
consists of the spin singlet and triplet channel:
\begin{eqnarray}
&&\{\hat{\Gamma}^{\alpha\beta}_a(p,p')\}_{\gamma\delta}=\Gamma_s(p,p')
i(\sigma_{y})_{\alpha\beta}i(\sigma_{y})_{\gamma\delta} \nonumber \\
&&+\Gamma_t(p,p')(\vec{\hat{t}}_p\times\vec{n})\cdot (\vec{\sigma} 
i\sigma_{y})_{\alpha\beta}(\vec{\hat{t}}_{p'}
\times\vec{n})\cdot (\vec{\sigma}i\sigma_{y})_{\gamma\delta}.
\end{eqnarray}
The symmetries of $\Gamma_{s}(p,p')$ and $\Gamma_{t}(p,p')$
in the momentum space
are characterized by the same irreducible
representation, and as a result,
$\Delta_{s}(p)$ and $\Delta_{t}(p)$ in eq.(\ref{gap}) have the same
symmetry in the momentum space.
The realized superconducting state is the mixture of the spin singlet 
and triplet states.~\cite{ede1,gor}
In this case, the possible pairing state is $s+p$ or $d+f$ or $g+h$,
and so forth. 

\section{A comment on the superconducting state realized in
${\rm CePt_3Si}$ : possibility of an unusual coherence effect}

Here, we would like to make a brief remark about
the pairing state realized in CePt$_{3}$Si.
The NMR measurement carried out by Yogi et al. shows
the existence of the coherence peak of $1/T_1T$ just below $T_c$, 
indicating the full-gap state without nodes.~\cite{yogi}
On the other hand, the recent experiment on 
the thermal transport done by Izawa et al. supports
the existence of line nodes of the superconducting gap.~\cite{matsu}
A possible resolution of this discrepancy is that
the line nodes of the superconducting gap are 
generated accidentally at the magnetic Zone boundary
which emerges as a result of the antiferromagnetic phase 
transition at $T_N=2.2$ K, and crosses the Fermi surface.
For such accidental nodes without the sign change of
the superconducting gap function, 
the coherence factor of $1/T_1T$ does not vanish,
resulting in the enhancement of the coherence peak just below $T_c$.
In this case, a plausible candidate for the pairing state is  
the $s+p$ wave state.
An important point which we would like to stress here is that
even when the superconducting state is dominated by 
the $p$ wave pairing; i.e. $\Delta_s(p) \ll \Delta_t(p)$,
the coherence factor which enters into $1/T_1T$ does not vanish.
This contrasts with the case of 
the usual $p$ wave state realized in centrosymmetric superconductors,
where the coherence factor of $1/T_1T$ disappears.
This is understood as follows.
For simplicity, we ignore electron correlation effects.
Then, in noncentrosymmetric superconductors, 
the nuclear relaxation rate is given by,~\cite{sch}
\begin{eqnarray}
&&\frac{1}{T_1T}\propto  \nonumber \\
&&\lim_{\omega\rightarrow 0}
\frac{1}{\omega}{\rm Im}\Bigl[T\sum_{\varepsilon_m}\sum_{p,p'}
\{{\rm Tr}[\frac{\sigma^{+}}{2}\hat{G}(p,\varepsilon_m+\omega_n)
\frac{\sigma^{-}}{2}\hat{G}(p',\varepsilon_m)] \nonumber \\
&&-{\rm Tr}[\frac{\sigma^{+}}{2}
\hat{F}(p,\varepsilon_m+\omega_n)
\frac{\sigma^{-}}{2}\hat{F}(p',\varepsilon_m)]\}
|_{i\omega_n\rightarrow \omega+i\delta}\Bigr] \nonumber \\
&&=\int \frac{d\varepsilon}{2\pi}\frac{1}{2T\cosh^2\frac{\varepsilon}{2T}}
\{[N_n(\varepsilon)]^2+[N_a(\varepsilon)^2]\}, \label{t1}
\end{eqnarray}
with $N_n(\varepsilon)$ and $N_a(\varepsilon)$ defined by the retarded Green's
functions as,
\begin{eqnarray}
N_n(\varepsilon)=-\sum_p\sum_{\tau=\pm}{\rm Im}
G^R_{\tau}(p,\varepsilon),
\end{eqnarray}
\begin{eqnarray}
N_a(\varepsilon)=-\sum_p\sum_{\tau=\pm}{\rm Im}
F^R_{\tau}(p,\varepsilon).
\end{eqnarray}
The expression of $1/T_1T$ (\ref{t1})
does not rely on the phase factor $\vec{\hat{t}}_p\times \vec{n}$
of the triplet component of the gap function (\ref{gap}).
The second term of the right-hand side of (\ref{t1}) gives
the non-zero contribution from the coherence factor, 
as in the case of conventional $s$ wave superconductors.
This property enhances the coherence peak of $1/T_1T$
prominently.
It may be important to take into account this unusual coherence effect
for clarification of the origin of the notable coherence peak of $1/T_1T$
in CePt$_3$Si.~\cite{yogi}
It is also intriguing to explore the unusual 
coherence effect on other response functions, 
such as the ultrasonic attenuation.
We would like to address this issue elsewhere.

\begin{figure}[h]
\includegraphics*[width=4cm]{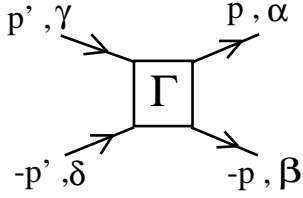}
\caption{Four point vertex in the particle-particle channel
}
\end{figure}

\section{Many-body effects on the magnetoelectric transport 
in the superconducting state}

In this section, we consider electron correlation effects on 
the magnetoelectric transport
in the superconducting state first found by Edelstein and later
discussed in detail by Yip; i.e. the emergence of the paramagnetic 
supercurrent induced by the Zeeman magnetic field in the direction
normal to the $\vec{n}$ vector.~\cite{ede1,yip}

We consider the charge current flowing in the $y$ direction which 
is defined as,
\begin{eqnarray}
J_y=T\sum_{\varepsilon_n,p}\frac{1}{2}{\rm Tr}[\hat{V}_{0y}\mathcal{G}(p)]
\end{eqnarray}
with
\begin{equation}
\hat{V}_{0y}=
\left(
\begin{array}{cc}
\hat{v}_{0y}(\vec{p}) & 0 \\
0 & -\hat{v}_{0y}^{t}(-\vec{p})
\end{array}
\right)
\end{equation}
and 
$\hat{v}_{0y}=\partial_{p_y}\varepsilon_p+\alpha(\vec{n}\times \vec{\sigma})$.

The transport coefficient which characterizes the Edelstein's 
magnetoelectric effect is given by,
\begin{eqnarray}
&&\mathcal{K}_{yx}\equiv \left.\frac{\partial J_y}{\partial H_x}\right|_{H_x=0}
\nonumber \\
&&=-T\sum_{n,p}\frac{1}{2}
{\rm Tr}[\hat{V}_{0y}\mathcal{G}(p)
\frac{\partial \mathcal{G}^{-1}(p)}
{\partial H_x}\mathcal{G}(p)]|_{H_x=0} \label{k1}
\end{eqnarray}

Following ref.3,
to simplify the expression of $\mathcal{K}_{yx}$
we use the Ward's identity for the current vertex,
\begin{eqnarray}
\frac{\partial\hat{G}(p)}{\partial p_y}&=&
\hat{G}(p)\hat{\tilde{V}}_y(\vec{p})\hat{G}(p)
+\hat{F}(p)\hat{\tilde{V}}^t_y(-\vec{p})
\hat{F}^{\dagger}(\vec{p},-i\varepsilon) \nonumber \\
&&+\hat{R}(p), \label{wa}
\end{eqnarray}
where $\hat{\tilde{V}}_y(\vec{p})=
\hat{v}_{0y}+\partial\hat{\Sigma}(p)/\partial p_y$, and
\begin{eqnarray}
\hat{R}(p)=
\left(
\begin{array}{cc}
 0 & r(p) \\
 r^{*}(\vec{p},-i\varepsilon) & 0
\end{array}
\right),
\end{eqnarray} 
\begin{eqnarray}
r(p)=\left[\frac{G_{+}-G_{-}}{2\alpha|\vec{t}_p|}
-G_{+}G_{-}+F_{+}F_{-}\right]
\hat{t}_{+}\Lambda_{+-}^{cy}(p),
\end{eqnarray}
\begin{eqnarray}
\Lambda_{+-}^{cy}(p)=\hat{t}_x
(\alpha+\frac{\partial {\rm Re}\Sigma_{\uparrow\downarrow}}{\partial p_y})
-\hat{t}_y
\frac{\partial {\rm Im}\Sigma_{\uparrow\downarrow}}{\partial p_y}.
\end{eqnarray}
Then, from (\ref{g1}), (\ref{k1}), and (\ref{wa}), we obtain,
\begin{eqnarray}
\mathcal{K}_{yx}&=&
-2T\sum_{n,p}{\rm Tr}\left[\hat{F}(p)\hat{v}_{0y}^t(-\vec{p})
\hat{F}^{\dagger}(p)\left(\mu_{\rm B}\sigma_x-
\frac{\partial \hat{\Sigma}}{\partial H_x}\right)\right]  \nonumber \\
&-&2\mu_{\rm B}\alpha T\sum_{n,p}F_{+}F_{-}\hat{t}_x
\Lambda_{+-}^{sx}(p) \nonumber \\
&+&T\sum_{n,p}{\rm Tr}\left[\frac{\partial \hat{\Sigma}}{\partial p_y}
\frac{\partial \hat{G}}{\partial H_x}\right]
-T\sum_{n,p}{\rm Tr}\left[\frac{\partial \hat{\Sigma}}{\partial H_x}
\frac{\partial \hat{G}}{\partial p_y}\right], \label{sc}
\end{eqnarray}
where the three-point vertex function is,
\begin{eqnarray}
\Lambda_{+-}^{sx}(p)=\hat{t}_x(1-\frac{1}{\mu_{\rm B}}\frac{\partial {\rm Re}
\Sigma_{\uparrow\downarrow}}{\partial H_x})+\frac{\hat{t}_y}{\mu_{\rm B}}
\frac{\partial {\rm Im}
\Sigma_{\uparrow\downarrow}}{\partial H_x}.
\end{eqnarray}
The last two terms of (\ref{sc}), which emerge as a result of 
Fermi liquid corrections, can be rewritten by using the Luttinger-Ward's
identity generalized to the superconducting state.
The Luttinger-Ward's identity in the normal state reads,~\cite{lut,lut2}
\begin{eqnarray}
T\sum_{n,p}{\rm Tr}[\hat{\Sigma}\frac{\partial \hat{G}}{\partial p_{\mu}}]=0.
\end{eqnarray}
This relation is obtained by differentiating all closed linked
diagrams with respect to $p_{\mu}$.~\cite{lut,lut2}
In the superconducting state, a similar analysis leads, 
\begin{eqnarray}
T\sum_{n,p}{\rm Tr}[\hat{\Sigma}\frac{\partial \hat{G}}{\partial p_{\mu}}]
&+&T\sum_{n,p}{\rm Tr}[\hat{\Delta}^{\dagger}\frac{\partial \hat{F}}
{\partial p_{\mu}}] \nonumber \\
&+&T\sum_{n,p}{\rm Tr}[\hat{\Delta}\frac{\partial \hat{F}^{\dagger}}
{\partial p_{\mu}}]=0. \label{lw}
\end{eqnarray}
Differentiating eq.(\ref{lw}) with respect to $H_x$, and
integrating over $p_y$ by parts, we have,
\begin{eqnarray}
&&T\sum_{n,p}{\rm Tr}\left[\frac{\partial \hat{\Sigma}}{\partial p_y}
\frac{\partial \hat{G}}{\partial H_x}\right]
-T\sum_{n,p}{\rm Tr}\left[\frac{\partial \hat{\Sigma}}{\partial H_x}
\frac{\partial \hat{G}}{\partial p_y}\right] \nonumber \\
&&=
-T\sum_{n,p}{\rm Tr}\left[\frac{\partial \hat{\Delta}^{\dagger}}{\partial p_y}
\frac{\partial \hat{F}}{\partial H_x}\right]
+T\sum_{n,p}{\rm Tr}\left[\frac{\partial \hat{\Delta}^{\dagger}}{\partial H_x}
\frac{\partial \hat{F}}{\partial p_y}\right] \nonumber \\
&&-T\sum_{n,p}{\rm Tr}\left[\frac{\partial \hat{\Delta}}{\partial p_y}
\frac{\partial \hat{F}^{\dagger}}{\partial H_x}\right]
+T\sum_{n,p}{\rm Tr}\left[\frac{\partial \hat{\Delta}}{\partial H_x}
\frac{\partial \hat{F}^{\dagger}}{\partial p_y}\right]. \label{lw2}
\end{eqnarray}
We see from eqs.(\ref{sc}) and (\ref{lw2}) that
$\mathcal{K}_{yx}$ vanishes exactly in the normal state, 
as is consistent with 
the thermodynamic argument that a static magnetic field can not induce
non-equilibrium current flows.~\cite{ede1,yip}
The right-hand side of eq.(\ref{lw2}) consists of the terms which have
the form $\sum_{n,p} G_{\tau}F_{\tau'}$. 
The ratio of the contributions from these terms to those from
other terms of eq.(\ref{sc}) is of order $\Delta/E_F$, and thus
we can neglect the last two terms of (\ref{sc}) approximately.
Then, the magnetoelectric coefficient is expressed as,
\begin{eqnarray}
&&\frac{\mathcal{K}_{yx}}{e\mu_{\rm B}}= \nonumber \\
&&\sum_p\sum_{\tau=\pm 1}\tau v_{0y\tau}
\frac{z_{p\tau}^2\Delta_{p\tau}^2}{E_{p\tau}^{2}}
\left[\frac{{\rm ch}^{-2}\frac{E_{p\tau}}{2T}}{2T}-
\frac{{\rm th}\frac{E_{p\tau}}{2T}}{E_{p\tau}}\right]
\Lambda^{sx}_{\tau}(p) 
\nonumber \\
&&+2\alpha\sum_p\frac{z_{p+}z_{p-}\Delta_{p+}\Delta_{p-}}
{E_{p+}^2-E_{p-}^2}\left[\frac{{\rm th}\frac{E_{p+}}{2T}}{E_{p+}}
-\frac{{\rm th}\frac{E_{p-}}{2T}}{E_{p-}}\right]
\hat{t}_x\Lambda^{sx}_{+-}(p) \label{kyx}
\end{eqnarray}
where
\begin{eqnarray}
\Lambda^{sx}_{\tau}(p)=\hat{t}_y(1-\frac{1}{\mu_{\rm B}}\frac{\partial {\rm Re}
\Sigma_{\uparrow\downarrow}}{\partial H_x})-\frac{\hat{t}_x}{\mu_{\rm B}}
\frac{\partial {\rm Im}
\Sigma_{\uparrow\downarrow}}{\partial H_x}
-\frac{\tau}{\mu_{\rm B}}\frac{\partial {\rm Re}\Sigma}{\partial H_x}.
\end{eqnarray}
It is noted that the vertex corrections due to electron correlation,
$\Lambda^{sx}(p)$,
which appear in the above expression (\ref{kyx}) 
is nothing but the vertex corrections to
the uniform spin susceptibility,
\begin{eqnarray}
&&\chi_{xx}= \nonumber \\
&&\mu_{\rm B}^2\sum_p\sum_{\tau=\pm 1}
\frac{z_{p\tau}}{2E_{p\tau}^{2}}
\left[\frac{\varepsilon_{p\tau}^{*2}{\rm ch}^{-2}\frac{E_{p\tau}}{2T}}{2T}-
\frac{\Delta_{p\tau}^2{\rm th}\frac{E_{p\tau}}{2T}}{E_{p\tau}}\right]
\hat{t}_y\Lambda^{sx}_{\tau}(p) \nonumber \\
&&+ \mu_{\rm B}^2\sum_p\left[
\frac{\varepsilon_{p+}^{*}{\rm th}\frac{E_{p+}}{2T}}{2E_{p+}}
-\frac{\varepsilon_{p-}^{*}{\rm th}\frac{E_{p-}}{2T}}{2E_{p-}}
\right]
\frac{\hat{t}_x}{\alpha|\vec{t}_p|}\Lambda^{sx}_{+-}(p). \label{chi}
\end{eqnarray}
Eq.(\ref{chi}) is easily obtained by differentiating
the $x$ component of the total magnetization 
$S^x=\mu_{\rm B}T\sum_{n,p}{\rm Tr}[\sigma_{x}\hat{G}(p)]$ with respect
to $H_x$. Note that the first term of the right-hand side of (\ref{chi})
is the Pauli paramagnetic contribution and the second one is 
the ``van Vleck'' term which arises from excitations between
spin-orbit split two bands.
Generally, in heavy-fermion systems, 
the magnitude of the uniform spin susceptibility is 
enhanced by the vertex corrections $\Lambda^{sx}(p)$.
In typical heavy-fermion systems including
CePt$_3$Si, the Wilson ratio 
$R_W=T\chi/C/(T\chi_0/C_0)\sim 2$, which implies that
the vertex corrections $\Lambda^{sx}$ is approximately of order 
the mass enhancement factor $1/z_{p\tau}$.~\cite{take}
Therefore, in eq.(\ref{kyx}) effects of the vertex corrections
and the mass renormalization factors $z_{p\tau}$ 
cancel with each other.
This cancellation holds as long as there is no strong ferromagnetic
spin fluctuation which increases notably 
the magnitudes of the vertex corrections $\Lambda^{sx}$.
Another important feature of eq.(\ref{kyx}) is the absence of
the backflow term of the charge current, which usually exists in
the non-equilibrium current flow. (See the discussion on the usual 
Meissner current in the next section.)
This is related to the fact that the current induced by a static magnetic field
is a dissipationless equilibrium flow.
As a result, {\it the Fermi liquid corrections do not exist
in this magnetoelectric coefficient for heavy fermion superconductors,
provided that there is no ferromagnetic fluctuation.}
This is one of the main result of this paper.
In terms of the Kubo formula, 
the absence of electron correlation effects for $\mathcal{K}_{yx}$
is understood as follows.
$\mathcal{K}_{yx}$ is given by the correlation function
of the current and spin density operators.
The spin density vertex is renormalized by electron correlation in
the opposite way to the current vertex, resulting in the cancellation
of the mass renormalization factors.
The important implication of this result is that
the Zeeman-field-induced paramagnetic supercurrent is not suppressed by
electron correlation effects in contrast with
the usual diamagnetic supercurrent of which the magnitude is much
reduced by the large mass enhancement in heavy fermion systems.
This property may make the experimental observation of the magnetoelectric 
effect easier, as discussed in the next section.

\section{Implications for experimental observations}

On the basis of the formula (\ref{kyx}), we would like to
discuss how the Zeeman-field-induced paramagnetic supercurrent
is experimentally observed. 

When the magnetic field is applied in the $x$ direction,
the London equation is modified to,
\begin{eqnarray}
\vec{J}_s=-\frac{c}{4\pi \lambda^2}\vec{A}+\mathcal{K}_{yx}(\vec{n}\times
\vec{H}_x), \label{lon}
\end{eqnarray}
where $\vec{A}$ is the vector potential.
Since the applied magnetic field always induces both the diamagnetic 
and the paramagnetic supercurrent in the system 
with the Rashba spin-orbit interaction, 
it is important for the experimental observation of this effect 
to discriminate between these two supercurrents.
If one measures currents simply attaching leads to the sample and applying 
an in-plane magnetic field, the Zeeman-field-induced
paramagnetic supercurrent can not flow in the sample because it generates
the Joule heat in the normal metal leads.~\cite{com} In this case, 
the magnetoelectric effect is canceled with 
the non-zero phase gradient of the superconducting order parameter 
in the equilibrium state.

Here, to highlight the observation of the paramagnetic supercurrent, 
we consider an experimental setup composed of two superconducting samples
joined together as depicted in FIG.2.
The $\vec{n}$ vectors of the sample A and B are, respectively, 
given by $\vec{n}_{\rm A}=(0,0,1)$, 
and $\vec{n}_{\rm B}=-\vec{n}_{\rm A}$.  
The joined surface at the junction is normal to the $\vec{n}$ vectors.
The applied magnetic field in the $x$ direction 
$\vec{H}=(H,0,0)$ gives rise to 
the paramagnetic supercurrent in the sample A (B) in the direction 
$\vec{n}_{\rm A}\times \vec{H}$ ($-\vec{n}_{\rm A}\times \vec{H}$).
Then, electrons accumulated in the right (left) edge of the sample A (B)
transfer to the right (left) edge of the sample B (A)
to decrease the chemical potential difference between the samples A and B,
resulting in supercurrent flows circulating in the system. 
The applied magnetic field also induces the Meissner diamagnetic supercurrent.
As explained below, the paramagnetic contribution can be discriminated from 
the diamagnetic current by using the Volovik effect.~\cite{vol}

Let us first consider 
the coefficient of the diamagnetic Meissner supercurrent 
with the Fermi liquid corrections at zero temperature,
which is equal to the Drude weight in the normal state,~\cite{leg}
\begin{eqnarray} 
\frac{c}{4\pi \lambda^2}=\frac{e^2}{c}
\sum_{p,\tau=\pm 1}v_{p\tau}^{\mu *}J_{p\tau}^{\mu *}
\delta(\mu-\varepsilon_{p\tau}^{*}), \label{pen}
\end{eqnarray}
where the quasi-particle velocity is 
$v_{p\tau}^{\mu *}=\partial \varepsilon_{p\tau}^{*}/\partial p_{\mu}$, and
the charge current is
\begin{eqnarray}
J_{p\tau}^{\mu *}=v_{p\tau}^{\mu *}
+\sum_pf_{p\tau,p'\tau'}\delta (\mu-\varepsilon^{*}_{p\tau}) 
v_{p'\tau'}^{\mu *}.
\label{cur}
\end{eqnarray}
Here $f_{p\tau,p'\tau'}$ is the interaction between two quasiparticles.
The second term of (\ref{cur}) is the backflow term.
In heavy fermion systems, the mass renormalization factor $z_{p\tau}$ 
and the backflow term in the current $J_{p\tau}^{*}$ give rise to
the pronounced reduction of the Meissner coefficient.
For example, in CePt$_3$Si, $z_{p\tau}$ is estimated 
as $\sim 1/100$.~\cite{bau,hari}
If we assume the spherical Fermi surface,
eq.(\ref{pen}) reduces to $(e^2/c)v^{*}_Fn_s/p_F$.
Here $v^{*}_F$ is the renormalized Fermi velocity, and
$n_s$ the superfluid density, $p_F$ the Fermi momentum.
It should be notified that $n_s$ is renormalized by the backflow effect
of $J_{p\tau}^{*}$, and is not equal to the carrier density even
at zero temperature.
In particular, for heavy fermion systems in which Umklapp scattering
is expected to be strong, $n_s$ is smaller than the carrier density.
In contrast to the diamagnetic Meissner current, 
the Zeeman-field-induced paramagnetic supercurrent is not influenced by 
the many-body effects described above, as discussed in the previous section.
This difference of electron correlation effects between the diamagnetic and
paramagnetic supercurrents can be utilized 
to detect the magnetoelectric effect.

\begin{figure}[h]
\includegraphics*[width=5cm]{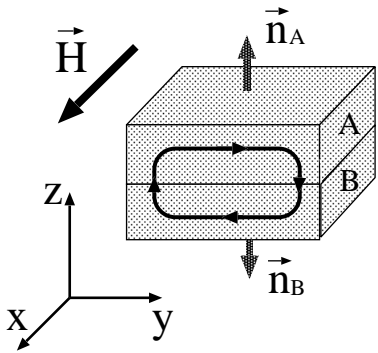}
\caption{An experimental setup for the detection of
the Zeeman-field-induced supercurrent. 
The $\vec{n}$ vectors of the two superconducting samples A and B
(depicted by the gray arrows)
are aligned in the directions $(0,0,1)$ and $(0,0,-1)$, respectively. 
The in-plane magnetic field $\vec{H}$ is applied in the $x$ direction.
The paramagnetic supercurrent circulates in the system as depicted by
the black thin arrows.   
}
\end{figure}

If the superconducting gap has a nodal structure 
as is often realized in some heavy-fermion superconductors, 
the existence of the paramagnetic supercurrent is indirectly 
observed through the Volovik effect on the single particle density of
states under the applied in-plane magnetic field for $H_{c1}<H<H_{c2}$.
Here $H_{c1}$ and $H_{c2}$ are, respectively, 
the lower and upper critical field.
In fact, the recent thermal transport measurements for CePt$_3$Si supports
the existence of the line node in the superconducting 
state of this system.~\cite{matsu}
Applying the semi-classical approximation based upon
the Doppler shift effect,~\cite{vol} and assuming
a spherical Fermi surface, we calculate the local density of states
from the modified London equation (\ref{lon}).
In the calculation of $\mathcal{K}_{yx}$, we use the fact that
the vertex corrections $\Lambda^{sx}_{\tau (+-)}$ is appropriately
approxiamted as $\sim z_{p\tau}^{-1}$ in typical heavy-ferimon systems
as discussed in Sec.IV., and 
expand eq.(\ref{kyx}) in terms of $\alpha p_F/E_F$ up to the lowest order.
The result at $T=0$ is
\begin{eqnarray}
\delta D_{\rm loc}(0)\sim \left|\sqrt{\frac{H}{H_{c2}}}
\frac{e^2v_{F}^{*}\Phi_0}{c\xi}
\pm \frac{H}{H_{c2}}\frac{e\mu_{\rm B}\alpha p_F\Phi_0n_0}{\pi\xi^2E_Fn_s}
\right|. 
\label{dos}
\end{eqnarray}
Here $\Phi_0=hc/(2e)$, and $n_0$ is the density of electrons.
$E_F$ is the unrenormalized Fermi energy.
The first term of the right-hand side of (\ref{dos}) is due to
the usual Volovik effect, and the second term linearly proportional to
the applied magnetic field stems from
the Zeeman-energy-induced paramagnetic supercurrent.
Thus, the magnetic-field-dependence distinguishes between the paramagnetic
and diamagnetic currents. 
The above behavior of the local density of states may be observed by
the measurement of the specific heat coefficient or
the thermal conductivity in sufficiently 
low field regions.~\cite{hir,fra,vec,cho}
In the above expression of $\delta D_{loc}(0)$, it is seen that
the conventional diamagnetic contribution is suppressed by 
the mass renormalization factor $z_{p\tau}$ which appears through $v^{*}_F$, 
whereas the magnetoelectric contribution is not affected by this correlation
effect. 
It is also noted that the carrier density which enters into the paramagnetic
term is not $n_s$ but equal to the electron density $n_0$.
This is due to the absence of the backflow term in the Zeeman-energy-induced
supercurrent. 
As mentioned before, $n_s$ is affected by the backflow term. 
For simplicity, we assume that $n_s\approx n_0$ for a while.

In the case of CePt$_3$Si, according to the measurement of $H_{c2}$,
the coherence length $\xi\sim8.1\times 10^{-7}$ cm.~\cite{bau}
It is a bit difficult to estimate 
the renormalized Fermi velocity from experimental measurements.
Bauer et al. obtained $v_F^{*}\sim 5.29\times 10^{5}$ cm/s
from the data of $dH_{c2}/dT$ and the specific heat coefficient, 
assuming a spherical Fermi surface.~\cite{bau}
This value of $v_F^{*}$ is almost of the same order as
that obtained by combining the unrenormalized 
Fermi velocity computed from the LDA method and
the mass enhancement factor $z^{-1}\sim 100$ estimated from the specific heat
measurement.~\cite{hari,bau}
According to the LDA band calculations,~\cite{hari,sam} 
the spin-orbit splitting is not so small compared to
the Fermi energy, and may be approximated as $\alpha p_F/E_F\sim 0.1$.
Then, for CePt$_3$Si, we have,
\begin{eqnarray}
\delta D_{\rm loc}(0)\sim \left|
1.0\times10^{-24}\sqrt{\frac{H}{H_{c2}}}\pm 0.48\times
10^{-24}\frac{H}{H_{c2}}\right|. \label{dos2}
\end{eqnarray}
It is remarkable that the contribution from the paramagnetic supercurrent 
(the second term of (\ref{dos2})) is
comparable to that from the Meissner supercurrent (the first term of 
(\ref{dos2})).
It should be stressed 
that the feasibility of the experimental observation
of the Edelstein's magnetoelectric effect is due to
the large mass enhancement in the heavy fermion system,
which suppresses strongly the Meissner supercurrent, but in contrast,
does not affect the Zeeman-field-induced paramagnetic supercurrent.
Moreover, if the superfluid density $n_s$ is reduced by
the backflow effect, the Meissner term of (\ref{dos}) is more suppressed
compared with the paramagnetic term, and thus
the observation of the magnetoelectric effect may become easier .

\section{Summary and Discussion}

We have investigated electron correlation effects on
the magnetoelectric transport phenomena 
in superconductors without inversion symmetry.
It is found that, in contrast to the Meissner diamagnetic supercurrent
which is much reduced by the mass enhancement factor in the absence of
translational symmetry, 
the Zeeman-field-induced paramagnetic supercurrent 
is not affected by the strong electron correlation
provided that ferromagnetic fluctuation is not developed.
Because of this remarkable property,
the experimental detection of the magnetoelectric effect may be
more feasible in heavy-fermion superconductors without inversion symmetry such
as CePt$_3$Si, where the enormous mass enhancement suppresses
the magnitude of the Meissner supercurrent, than in
conventional metals with moderate effective electron mass.
We have proposed the experimental setup for the observation of
the magnetoelectric effect in CePt$_3$Si which utilizes the Volovik effect.
It has been also pointed out that in noncentrosymmetric $p$ wave 
superconductors, the coherence effect on the nuclear relaxation rate
$1/T_1T$ is similar to that of conventional $s$ wave superconductors.

Finally, we would like to comment on the implication of our results
for UIr, which is the recently discovered ferromagnetic superconductor
without inversion symmetry.~\cite{uir}
UIr exhibits superconductivity under high pressure
in the vicinity of the phase boundary between ferromagnetic 
and non-magnetic states.
The resistivity of this system increases remarkably as the applied pressure
approaches to the critical value at which the ferromagnetism disappears, 
indicating the existence of ferromagnetic critical fluctuation.
In this case, the magnetoelectric coefficient eq.(\ref{kyx})
may be enhanced by the three-point vertex functions $\Lambda^{sx}_{\tau(+-)}$,
of which the magnitudes are much increased by ferromagnetic fluctuation,
provided that the spin easy axis is taken as the $x$ axis.
Since, the crystal structure of UIr is monoclinic, and does not
possess any mirror planes, 
the Rashba spin-orbit interaction with the $\vec{n}$ vector perpendicular to
the spin easy axis should always exists.
Thus, the magnetoelectric effect strongly enhanced by
ferromagnetic fluctuation may be observed in UIr under
an applied magnetic field parallel to the spin easy axis.

\acknowledgments{}

The author would like to thank K. Yamada, 
Y. Matsuda, and H. Ikeda for invaluable discussions.
This work was partly supported by a Grant-in-Aid from the Ministry
of Education, Science, Sports and Culture, 
Japan.




\begin{thebibliography}{99}
\bibitem{com} If the attached leads are made of superconductors
with the transition temperature higher than that of the sample and
the same pairing symmetry as that of $\Delta_s(p)$ and $\Delta_t(p)$,
the Zeeman-field-induced supercurrent can flow in the system. 
Note that the tunneling amplitude between the leads 
and the noncentrosymmetric superconductor does not depend on
the symmetric properties of the $\vec{d}$ 
vector of the gap function (\ref{gap}),
$\vec{\hat{t}}_p\times\vec{n}$, 
as pointed out by I. A. Sergienko and S. H. Curnoe, in ref. 18.
The observation of the magnetoelectric effect 
may be possible also in this case, as discussed in Sec.V.
\end{thebibliography}
\end{document}